\documentclass[aps,prl,preprint,groupedaddress]{revtex4-1}
\usepackage{graphicx}

\begin{document}

\title{Ge$_{136}$ type-II clathrate as precursor for the synthesis of metastable germanium polymorphs: a computational study}

\author{Daniele Selli$^{1}$, Igor A. Baburin$^{2}$, Roman Marto\v{n}\'{a}k$^3$, Stefano Leoni$^{1}$}
\affiliation{$^1$Cardiff University, School of Chemistry, Park Place, CF10 3AT, Cardiff, UK\\
$^2$Technische Universit\"at Dresden, Institut f\"ur Physikalische Chemie, 01062 Dresden, Germany\\
$^3$Department of Experimental Physics, Comenius University, Mlynsk\'{a} Dolina F2, 842 48 Bratislava, Slovakia}


\begin{abstract}
\bf{The response to compression of the clathrate type-II structure Ge(cF136) is investigated by means of \textit{ab initio} small-cell metadynamics at different temperatures  and pressures. At lower pressure $p$=2.5 GPa the metastable metallic bct-5 phase competes against $\beta$-Sn Ge(tI4), which forms at higher pressures. On lowering temperature, the presence of amorphous intermediates obtained from Ge$_{136}$ is more pronounced and is instrumental to the formation of denser structural motifs, from which metallic bct-5 can form. Therein, anisotropic box fluctuations promote phase formation. The metadynamics runs are analysed in depth using a set of topological descriptors, including coordination sequence and ring statistics. Differences in the structural landscape history and amorphous intermediates are critical for the selective formation of particular metastable polymorphs, towards turning crystal structure predictions into actual materials.}
\end{abstract}

\maketitle

\section{Introduction}

Group-IVa elements (tetrels) display an extended polymorphism, reflected in a wide range of properties. Due to their technological relevance, semiconducting silicon and germanium have motivated repeated investigations on their polymorphism~\cite{Mujica:2003tb, Cui:2009kd, Chen:2011bx, Schwarz:2004fs, Katzke:2007cn, Mujica:2015eo}. Germanium displays higher carrier mobility than silicon and finer band gap tunability~\cite{Claeys:2010ij}. In Ge lowering of phonon frequencies promotes electron-phonon coupling towards superconductivity~\cite{Chen:2011bx,Cui:2009kd}. Metallization occurs in silicon and germanium upon compression~\cite{Mujica:2003tb}. The possibility of metallic germanium under room conditions is very intriguing and intensively debated~\cite{Cui:2009kd,Li:2010ge}, while pressure favours superconductivity in elemental Ge~\cite{Chen:2011bx}. 

Germanium bears similarities with silicon~\cite{Mujica:2003tb} by comparatively higher transition pressures~\cite{Lewis:1994tt}. Upon compression semiconducting Ge (cF8) transforms into $\beta$-tin type (tI4, space group I4$_1$/amd) at about 10 GPa~\cite{Menoni:1986wv}, and then to Imma phase~\cite{Nelmes:1996ip}, simple hexagonal (P6/mmm)~\cite{Vohra:1986ge}, followed by an orthorhombic Cmca phase~\cite{Takemura:2000tc} and finally upon further compression above 180 GPa, by the hexagonal close-packed arrangement (P6$_3$/mmc)~\cite{Takemura:2000tc}. Additional metastable polymorphs can be expected, if high-pressure experiments are designed to influence nucleation, by changed decompression protocols, low temperatures~\cite{Brazhkin:1995gy}, non hydrostaticity~\cite{Haberl:2014ie, Schwarz:2008wl}, or by choosing a different Ge allotrope as starting material~\cite{wosylus:2008pshs}. 

Open framework structures like clathrates~\cite{Cros:1970gy,Kasper:1965kp,SanMiguel:1999ev} are enjoying a renewed interest for they may host distinct electronic, magnetic, spectral and transport properties~\cite{Melinon:1998fy,Demkov:1996gu,Saito:1995ff,Nesper:1993fga,OKeeffe:1992gf}. Type-II clathrate Ge(cF136) was synthesised from a salt precursor Na$_{12}$Ge$_{17}$, by mild oxidation with HCl. It is stable at room condition and subsists up to 693 K~\cite{Guloy:2006gi}. Other reported metastable modifications , Ge(tP12), Ge(cI16) ($\gamma$-silicon type, BC8), can be generated by decompression~\cite{Nelmes:1993va,Bates:1965fqa,Bundy:1963hy}, while Ge(hR8)~\cite{Schwarz:2008wl} results from direct compression of cF136. In all these allotropes, the four-bonded atoms adopt nearest-neighbour distances that are similar to those of diamond-type Ge(cF8).


In a previous work~\cite{Selli:2013coa}, we have used metadynamics to investigate structural transformations between diamond Ge(cF8) and Ge(tI4), $\beta$-Sn type structure. In order to systematically approach the study of phase transitions mechanisms, the use of molecular dynamics accelerated technique is mandatory to efficiently overcome high energy barriers. Metadynamics explores free energy surfaces by depositing an history-dependent bias along selected collective variables (CV). To investigate pressure-induced polymorphism the whole cell can act as CV. 

Along the diamond$\rightarrow$ $\beta$-tin transition, metadynamics visited an intermediate of bct-5 topology (I4/mmm).
This hypothetical five-connected structure is metallic, mechanically stable at room conditions, and superconducting at low temperature. In the metadynamics run, only one box parameter was markedly affected, suggesting non-hydrostatic shearing as the protocol of choice towards obtaining bct-5.

\begin{figure}
\begin{center}
\includegraphics[width=0.75\textwidth,keepaspectratio]{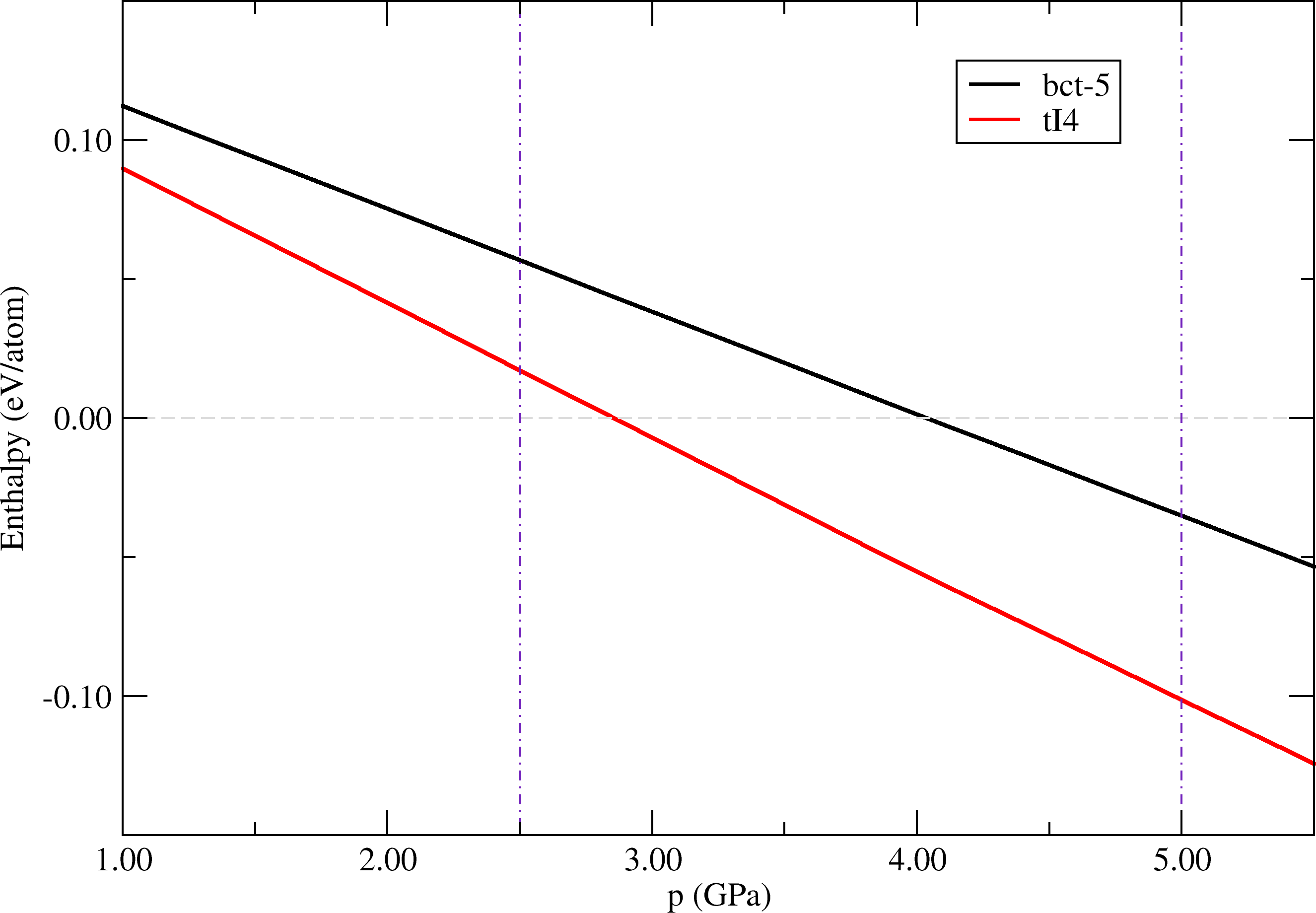}
\end{center}
\caption{Enthalpy differences between Ge$_{136}$ (cF136), $\beta$-tin (Ge(tI4)) and bct-5. Ge$_{136}$ represents the baseline. The pressure values considered in this work are indicated as vertical lines at 2.5 and 5.0 GPa.}
\label{Fig_0}
\end{figure}




Despite the noticeable differences in atomic volumes (24.5 and 28.3 \AA$^3$, respectively), the compressibility of Ge(cF136), as revealed by the bulk moduli B$_0$ = 76$\pm$6 GPa~\cite{Schwarz:2008wl}, is very close to the value of cubic diamond Ge(cF 8), B$_0$ = 75 GPa~\cite{Guinan:1974ej}. Total energy calculations (SIESTA~\cite{Soler:2002wq}) indicate that Ge(cF136) is the lowest-lying modification among metastable germaniums.


The transition pressure between cF136, cI4 and bct-5 were evaluated based on the enthalpy equivalence between phases, $E_1 + pV_1 = E_2 + pV_2$. The difficulty in accessing bct-5 from direct, isotropic compression lies in the presence and relative stability of $\beta$-tin. A strategy must therefore be designed, that takes into account non-hydrostatic compression (see above and ~\cite{Selli:2013coa}) of a suitable intermediate as learned from a recent~\cite{PhysRevB.105.134107} study. In the latter,  the pressure-induced amorphization and subsequent fast re-crystallization of Ge$_{136}$ into $\beta$-tin or low-density amorphous (LDA) Ge were investigated using \textit{ab initio} molecular dynamics and metadynamics. Metadynamics was used up to the occurrence of amorphous intermediates in the 136 atoms box, followed by plain equilibrium molecular dynamics protocols until phase crystallization. Fast recrystallization was also observed in very high-density amorphous (VHDA) Si~\cite{deringer_nature_2021} recently, likely owing to common tetrahedral structural motifs. 

In this work, differently from~\cite{PhysRevB.105.134107} we apply metadynamics beyond the stage of system amorphization to systematically identify configurations kinetically accessible from disordered intermediates. Two simulations pressures were therefore selected, p=5.0 GPa and p=2.5 GPa, above and below the threshold pressure for $\beta$-tin (Ge(tI4), Fig.~\ref{Fig_0}). All small-cell metadynamics runs were performed based on the 34 atoms primitive cell of Ge$_{136}$. The choice of a small simulation box shall favor anisotropic box fluctuations and also assist disordered-ordered system transitions as demonstrated by small-cell molecular dynamics applied to ice polymorphism ~\cite{Buch_JCP_123_2005}. 



\section{results}


\begin{figure}
\begin{center}
\includegraphics[width=0.75\textwidth,keepaspectratio]{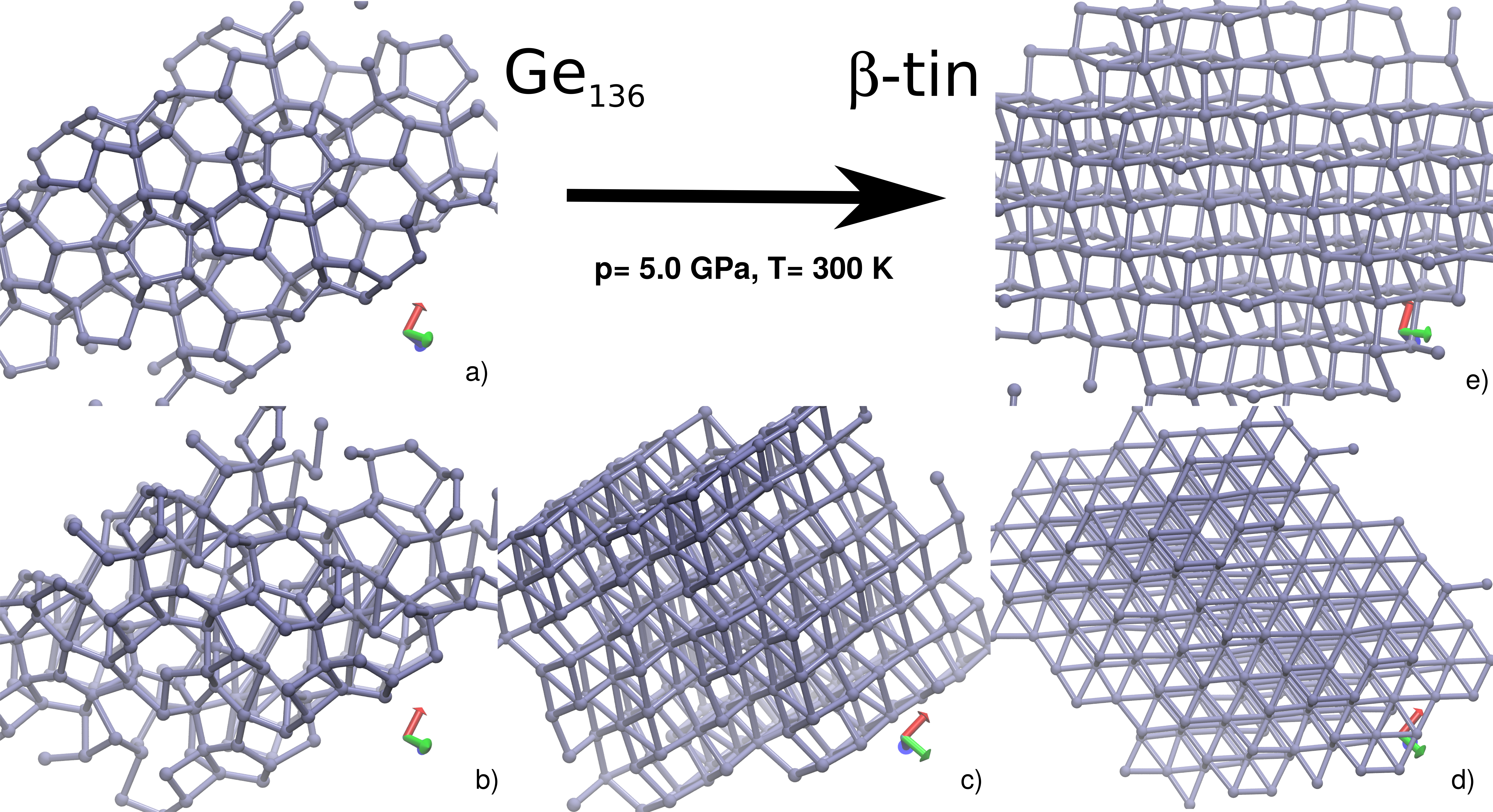}
\end{center}
\caption{Snapshots from a metadynamics trajectory at p=5.0 GPa and T=300 K. The corresponding timesteps are indicated in Fig.~\ref{Fig_2} as thick marks on the horizontal axis. a) Ge$_{136}$ transforms into e) $\beta$-tin (Ge(tI4)). The metatrajectory visits a number of intermediates (b, c, d) as shown, refer to text for details.} 
\label{Fig_1}
\end{figure}



\begin{figure}
\begin{center}
\includegraphics[width=0.75\textwidth,keepaspectratio]{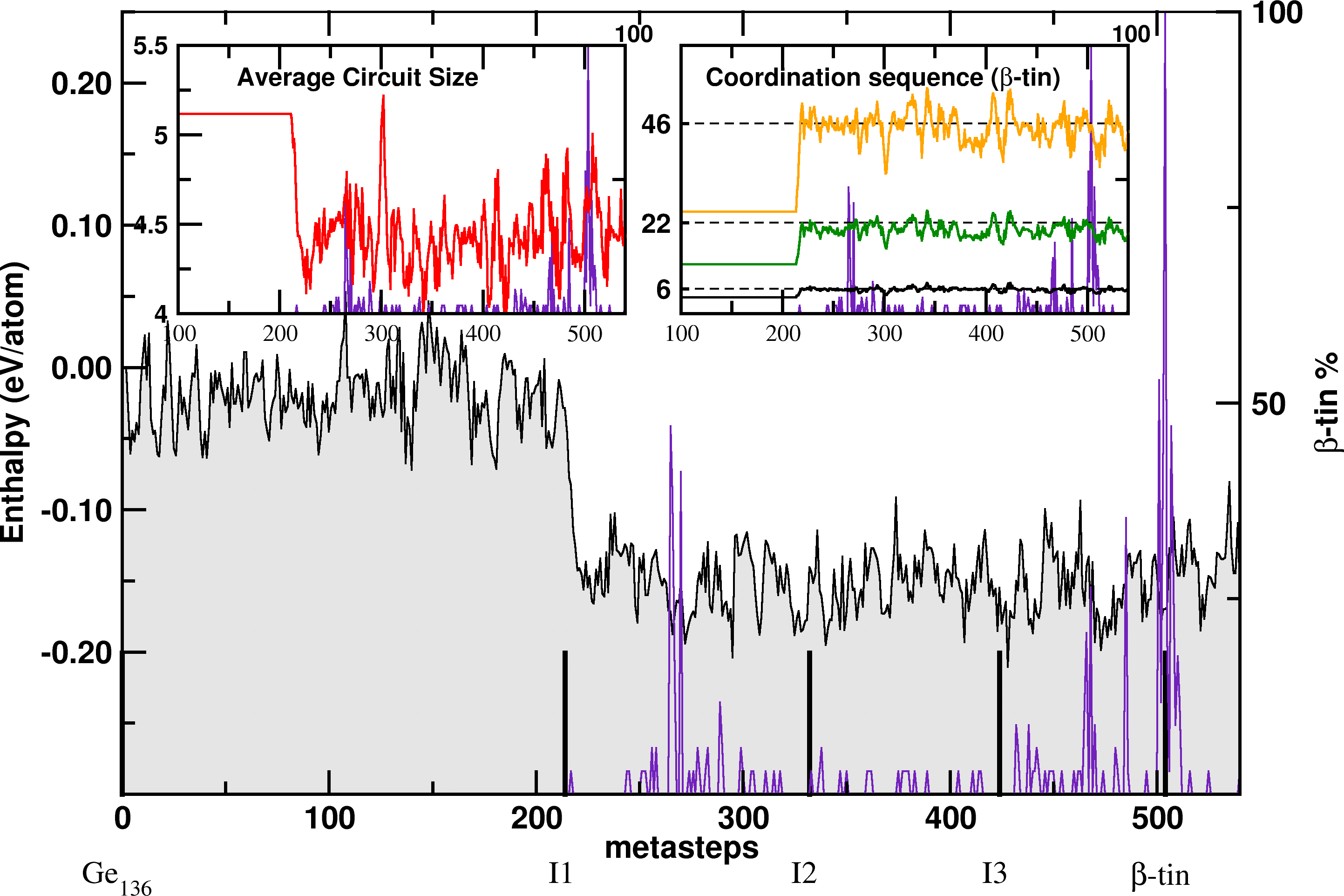}
\end{center}
\caption{Evolution of structure descriptors for the metadynamics trajectory of Fig.~\ref{Fig_1}: Average Circuit Size (small panel, left), Coordination Sequence (small panel, right), and Enthalpy/atom. Completion of the transformation of Ge$_{136}$ into $\beta$-tin is represented as indigo line. Thick marks on the horizontal axis correspond to the metasteps of the five snapshots represented in Fig.~\ref{Fig_1} a-e.} 
\label{Fig_2}
\end{figure}

In Fig.~\ref{Fig_1} five metatrajectory snapshots from type-II clathrate Ge(cF136) to Ge(tI4) \textbf{at p=5.0 GPa and T=300 K} are shown, while Fig.~\ref{Fig_2} summarises the evaluated descriptors for the same metatrajectory. The percentage of pure $\beta$-tin species is displayed on the opposite Y axis (indigo in Fig.~\ref{Fig_2}). The metasteps corresponding to the snapshots of Fig.~\ref{Fig_1} are indicated in Fig.~\ref{Fig_2} as thick marks on the horizontal axis, and labelled (Ge$_{136}$, I1, I2, I3, $\beta$-tin).

The stability of Ge(cF136) is reflected in the large number of metasteps spent lingering in the 
initial basin. Only after more than 100 ps (about 210 metasteps) a sudden enthalpy drop (Fig.~\ref{Fig_2}, main panel) is observed, as the system moves into a regime of amorphous, but rather
dense structures (Fig.~\ref{Fig_1} b). The average circuits size (ACS) and enthalpy
display in fact minima, while the coordination sequence (Fig.~\ref{Fig_2}, right inset) 
rapidly increases. Subsequently, the system remains of mixed structural character, as measured by the above indicators, with a single attempt of forming $\beta$-tin (indigo peaks) around metastep 265.

To the ACS peak between metastep 295 and 305 corresponds a marked drop in the coordination sequence. In this region thermal fluctuations
bring the system close to $\beta$-tin (Fig.~\ref{Fig_2}, $\beta$-tin \%), however the structure remains rather
expanded and does not fully lock-in. The configuration at metastep 332 (Fig.~\ref{Fig_1} c) is representative for the subsequent 40 metasteps, and contains mixed structural motifs, comprising remaining $\beta$-tin like features (4+2 coordination) and higher coordination numbers up to 12. 

Between metasteps 370 and 400 plateaux can be recognised by inspection of the ACS, CS and enthalpy graphs. This region features structures related
to $\alpha$-polonium type (cP1), but  distorted (while Ge is on the average sixfold coordinated, the angles of the octahedral coordination polyhedron deviate from 90$^{\circ}$) and more dense, as displayed by the coordination sequence average {6, 18, 42} against the expected {6, 18, 38}  for ideal $\alpha$-polonium (cP1). A similar occurrence of a denser structural motif is observed between steps 405 and 425, where a simple hexagonal structural pattern is closely matched (Fig.~\ref{Fig_1}, d), with a slightly expanded volume though, as it can be inferred from the coordination sequence {7.5, 25, 54} against {8, 26, 56} of ideal hP1. Cell fluctuations enheancements are characteristic of the metadynamics methods, and contribute to driving the systems to the final
$\beta$-tin basin, into which the whole system has transformed (Fig.~\ref{Fig_2}, $\beta$-tin \%; Fig.~\ref{Fig_1} e) after about 500 metasteps.

\begin{figure}
\begin{center}
\includegraphics[width=0.75\textwidth,keepaspectratio]{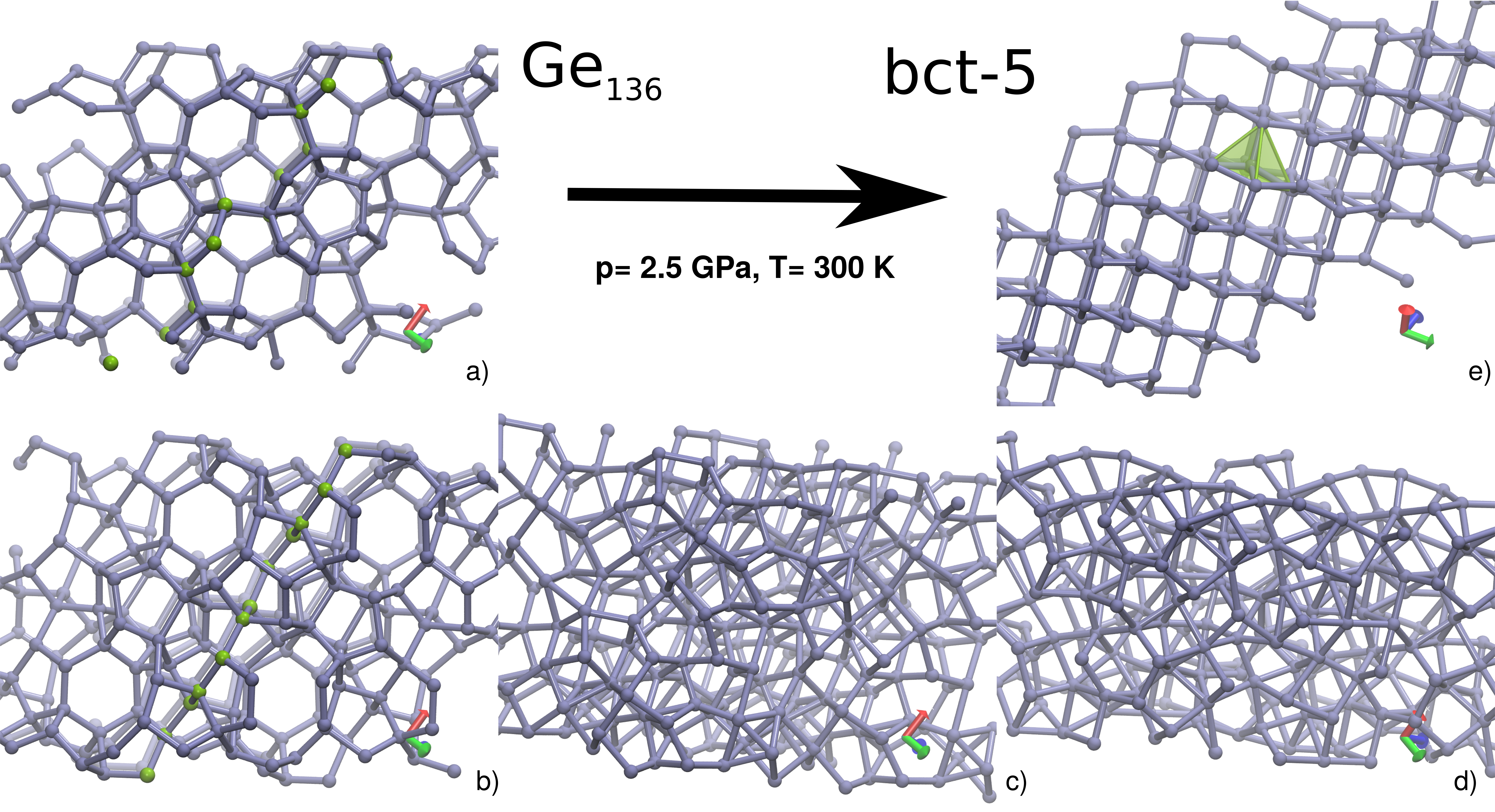}
\end{center}
\caption{Snapshots from a metadynamics trajectory at $p$=2.5 GPa and $T$=300 K. The corresponding timesteps are indicated in Fig.~\ref{Fig_4} as thick marks on the horizontal axis. a) Ge$_{136}$ transforms into e) bct-5. An unstable clathrate structure Ge(mC68) is visited, b). In a) and b) selected Ge atoms are highlighted in green to stress the closeness between the two structural motifs. The characteristic square-pyramidal coordination polyhedron of bct-5 is shaded in green in e).} 
\label{Fig_3}
\end{figure}

\begin{figure}
\begin{center}
\includegraphics[width=0.75\textwidth,keepaspectratio]{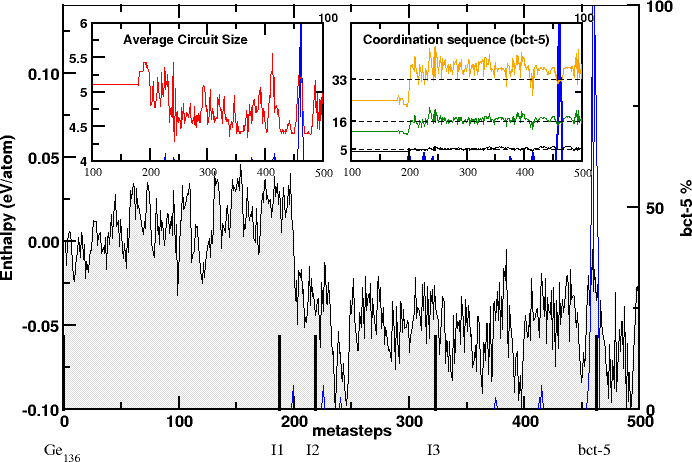}
\end{center}
\caption{Evolution of structure descriptors for the metadynamics trajectory of Fig.~\ref{Fig_3}: Average Circuit Size (small panel, left), Coordination Sequence (small panel, right), and Enthalpy/atom. Completion of the transformation of Ge$_{136}$ into bct-5 is represented as blue line. Thick marks on the horizontal axis correspond to the metasteps of the five snapshots represented in Fig.~\ref{Fig_3} a-e.}
\label{Fig_4}
\end{figure}

The same descriptors were applied to a metatrajectory \textbf{at $p$=2.5 GPa and $T$=300 K} that connects Ge(cF136) to bct-5. 
The percentage of pure bct-5 appears on the opposite Y axis, blue curve in Fig.~\ref{Fig_4}. 
The lower pressure regime markedly affects the meta-trajectory. After about 175 metasteps, Ge(cF136)  
distorts into an intermediate, which can be quenched (Fig.~\ref{Fig_3} b).

We characterised it with the Pearson symbol Ge(mC68). This clathrate-like structure is higher in energy than Ge(cf136)
($\Delta$E = E(cF8)$-$E(cF136) = 0.03 eV/atom, while $\Delta$E = E(cF8)$-$E(mC68) = 0.16 eV/atom), and can be relaxed in all parameters at $0 K$. 
The chosen projection of Fig.~\ref{Fig_3} a,b allows to appreciate the structural closeness of Ge(cF136) and Ge(mC68). 
A set of atoms has been highlighted (green spheres in Fig.~\ref{Fig_3} a,b) to illustrate the formation of chains as part of the transformation mechanism, which leaves large portions of Ge(cF136) unaffected. A short molecular dynamics run started from Ge(mC68) ($t$ = 10 ps, $p$ = 0 GPa, $T$ = 300 K) didn't show any appreciable structural deformation. Nonetheless, imaginary frequencies were found in the phonon dispersion
analysis (not shown), suggesting overall mechanical instability. In fact, after 25 metasteps, mC68 evolves into another
structural area. Here a single structural motif can hardly be extracted, for what can be described as intermediate configurations between
Ge(cF 136) and a denser, $\beta$-tin like phase: the average circuits size for this segment is $\sim$4.9,
intermediate between 5.1 for (cF136) and 4.67 for $\beta$-tin; the coordination sequence
has also increased with respect to Ge(cF136), indicating overall denser structures (Fig.~\ref{Fig_3} c), while also the enthalpy takes intermediate values. 

Thereafter, a long segment of the trajectory of about 175 metasteps consists of amorphous structures that occasionally show local order in correspondence to maxima in ACS and in CS, and to enthalpy minima. A representative snapshot is displayed in Fig.~\ref{Fig_3} d.

At metastep 430 the metadynamics trajectory reaches the basin of bct-5, as indicated by blue peaks in Fig.~\ref{Fig_4}. Its characteristic five-fold coordination is illustrated by a green pyramidal polyhedron, Fig.~\ref{Fig_3}e. bct-5 appears in a region of higher enthalpy values, in correspondence of a tiny local minimum. The bct-5 motif forms within a landscape dominated by low-crystalline motifs, from which $\beta$-tin crystallization is not observed.



Two relevant pressure protocols can therefore be identified. At relatively low pressure ($P$ = 2.5 GPa), a transition from Ge$_{136}$ to bct-5 can be observed, proceeding over amorphous intermediate steps. While higher pressure always favour the formation of Ge(tI4) from Ge(cF136), lower pressures admit bct-5 among the accessible configurations. Nonetheless, even in a pressure regime below its stability range like the one implemented here, Ge(tI4) remains a common transient motif linking four-fold to higher coordinations. This prompts for a thorough investigation of the factors that may determine the appearance of bct-5. 


\begin{figure}
\begin{center}
\includegraphics[width=0.75\textwidth,keepaspectratio]{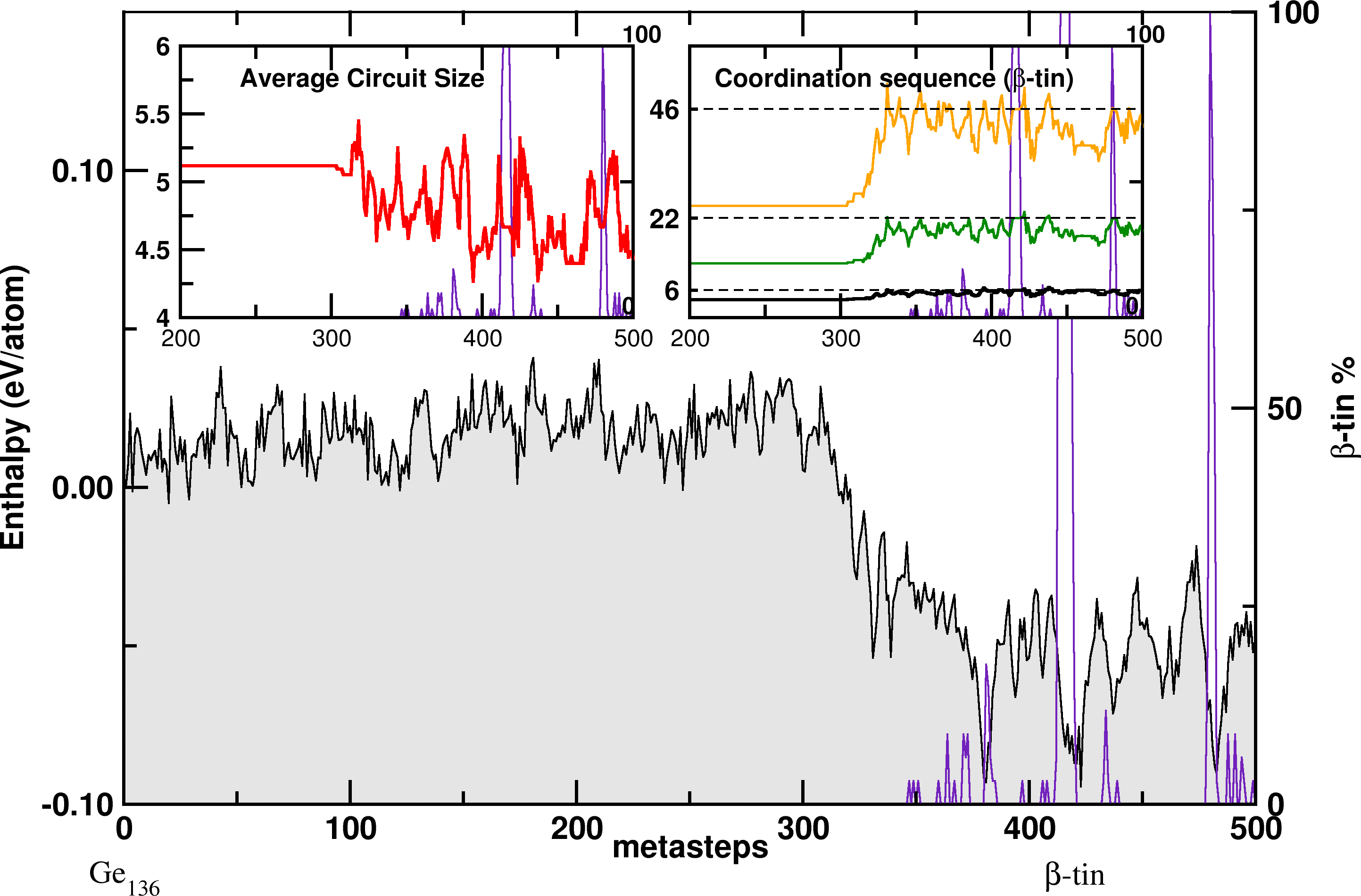}
\end{center}
\caption{Evolution of structure descriptors for a metadynamics trajectory ($p$=5.0 GPa, $T=$77 K): Average Circuit Size (small panel, left), Coordination Sequence (small panel, right), and Enthalpy/atom. Completion of the transformation of Ge$_{136}$ into $\beta$-tin is represented as indigo line.} 
\label{Fig_5}
\end{figure}

Metadynamics from Ge(cF136) was repeated \textbf{at 5.0 GPa and 77 K}, using the same protocol as for the previous runs. Lowering temperature caused the $\beta$-tin phase to sharply lock in, with a clear signature on the enthalpy profile, and remained well-defined over many metasteps, as shown in Fig.~\ref{Fig_5}. Also, the transition was markedly smoother as indicated by the gentle drop of the enthalpy profile, the overall progressive drop of the ACS and increase of CS. 

As the system escapes the Ge(cF136) well, it visits an open framework structure (metasteps 300-310), which is denser than the initial
structure and which displays a minimum in the ACS and an increased CS value (Fig.~\ref{Fig_5}, left and right insets). 
This is less of a metastable phase than an activated intermediate, which is  followed by a region of amorphous structures, as any coherent structural motif is hardly distinguishable, while neither ACS nor CS indicate the existence of a clearly defined intermediate crystalline structure.

Around metastep 380 the system re-crystallizes into a $\beta$-tin like structure (maximum in ACS and minima in CS values and enthalpy). Cell fluctuations
bring it back to an amorphous stage, which is the gateway to the exact $\beta$-tin basin. Between two 100\% $\beta$-tin peaks centred at metasteps 416 and 480, 
other amorphous phases are touched upon. Nonetheless, it can be expected that these effects can be further affected by re-tuning the choice of the Gaussian parameters in the simulation setup.

\begin{figure}
\begin{center}
\includegraphics[width=0.75\textwidth,keepaspectratio]{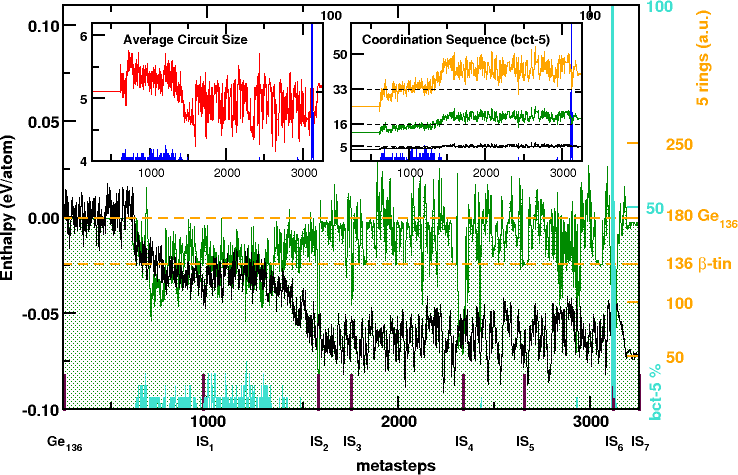}
\end{center}
\caption{Evolution of Average Circuit Size (small panel, left), Coordination Sequence (small panel, right), Enthalpy/atom and number of five-rings (gree line, main panel), for a metadynamics run at $p$=2.5 GPa and $T$=77 K. Eight thick marks on the horizontal axis indicate the metasteps relative to the eight snapshots of Fig.~\ref{Fig_7} a-h. On the opposite Y axis in the main panel, the characteristic number of five-rings for Ge$_{136}$ and $\beta$-tin are indicated, 180 and 136 respectively (orange labels and dashed orange lines). In bct-5, there are no five-rings. The progress of the  conversion from Ge$_{136}$ to bct-5 is indicated as a blue(insets)/turquoise line. }  
\label{Fig_6}
\end{figure}

The metadynamics \textbf{at p=2.5 GPa, T=77K} was performed with a slightly modified protocol, which entailed rescaling width and height of the Gaussian bias (see Section Methods for details). The rescaling to lower value would namely allow for a longer persistence in stable basins. A full analysis based on the same descriptors as for the other runs (enthalpy, ACS, CS), augmented by a statistics of 5-rings (green line) is presented in Fig.~\ref{Fig_6}, main panel. Five-fold rings introduce an additional means to sharply distinguish between Ge(cF136), Ge(tI4) and bct-5. Their characteristic 5-ring number is 180 and 136 respectively (orange on the opposite Y axis of the main panel graph in Fig.~\ref{Fig_6}), while there are no 5-rings in bct-5. 

In Ge(cF136) 5-rings occur as so-called K$_5$ graphs~\cite{hyde1999}, in which six 5-fold rings share each two edges around a Ge atom. They constitute therefore characteristic building blocks based on five-ring statistics, whose monitoring is therefore an excellent indicator of structure evolution.  

\begin{figure}
\begin{center} 
\includegraphics[width=0.75\textwidth,keepaspectratio]{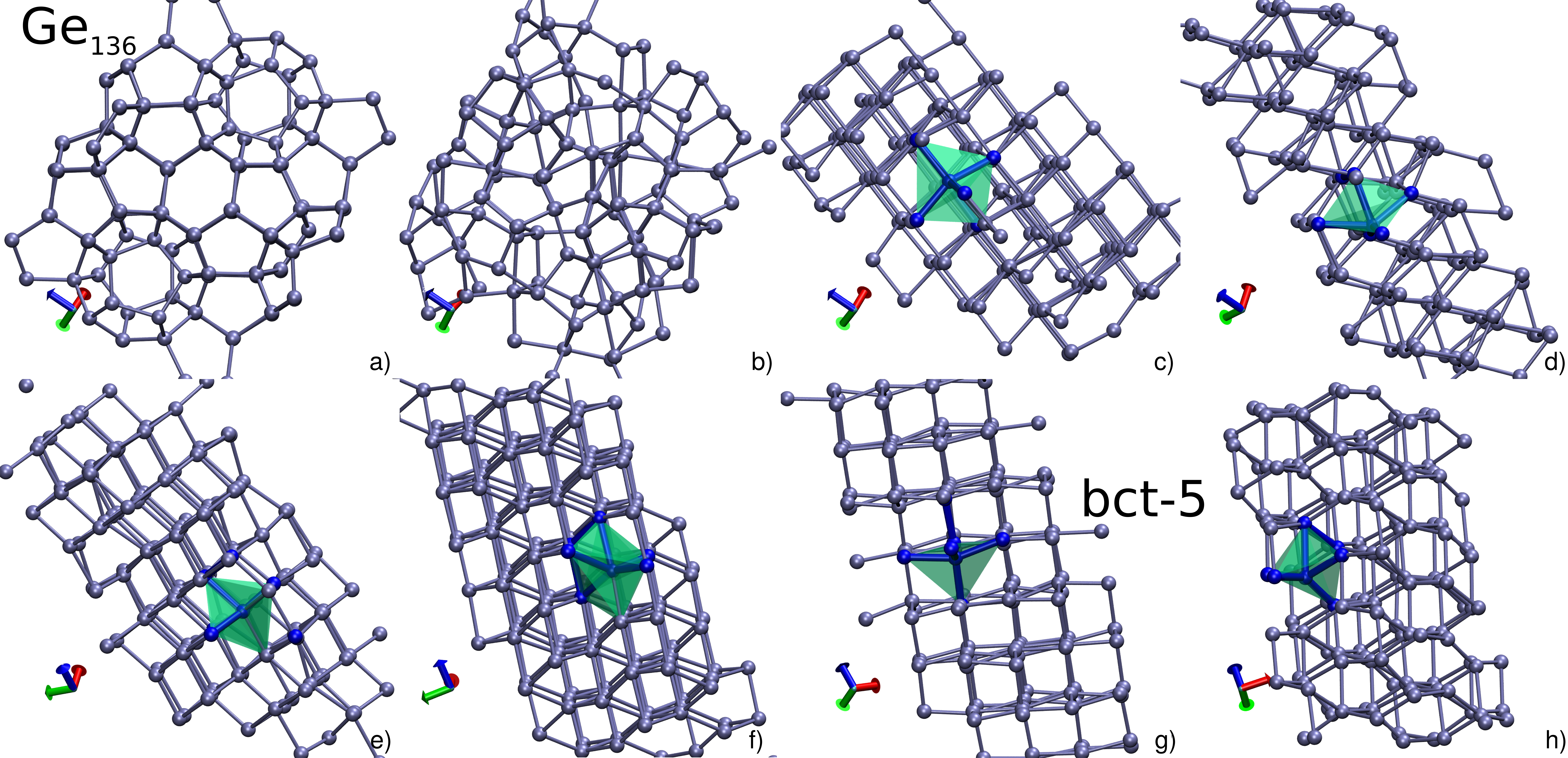}
\end{center}
\caption{Snapshots from a metadynamics trajectory at $p$=2.5 GPa and $T$=77 K. The corresponding timesteps are indicated in Fig.~\ref{Fig_6} as thick marks on the horizontal axis. a) Ge$_{136}$ transforms into g) bct-5 over several intermediates, b-f. The region of stability of Ge$_{136}$ a) is followed by a region of substantially amorphous motifs of intermediate enthalpy values b), which in turn evolves into a regime of lower enthalpy on the average c-h). The latter is characterised by the occurrence of local five-coordination. Therein, $\beta$-tin is occasionally visited, but fails to lock in at any point.}
\label{Fig_7}
\end{figure}

Considering major changes only, three regions can roughly be distinguished by enthalpy in Fig.~\ref{Fig_6}: a first segment from metastep 1 to 660, dominated by Ge(cF136), a second region between 660 and 1330, and a third region (1580-3260 metasteps) of overall lower enthalpy (including fluctuations). The CS indicates that the first two regions are  dominated by tetrahedral motifs (first CS=4, Fig.~\ref{Fig_6}, left inset, black line), while denser structures are found in the third region. Therein, the number of five-rings remains largely within (occasionally above) a well defined region (orange dashed line in Fig.~\ref{Fig_6}, main panel), while depletion of that figure occurs only episodically. In Fig.~\ref{Fig_7}, eight snapshots from the meta-trajectory are presented. Their corresponding metasteps are indicated on the main horizontal axis in Fig.~\ref{Fig_6} (vertical sticks labeled IS), typically in correspondence of a lowering of the number of 5-rings (Fig.~\ref{Fig_6}, green line). 

Ge$_{136}$ lasts till metastep 660 (Fig.~\ref{Fig_7}, a). A representative configuration of the second region would still contain 5-rings, which at this stage maintain partially fused 5-rings (Fig.~\ref{Fig_7}b, IS$_1$). After metastep 1330 structural motifs of higher 5-ring number are more prone to configuration fluctuations, i.e. they are rapidly evolving and do not leave any durable trace on any of the descriptors of interest, CS and enthalpy for instance. Correspondingly, configurations with fewer 5-rings are better characterised. At metastep 2430, to a spike-shaped depletion in the five-ring curve corresponds a structure, which displays features of bct-5 (Fig.~\ref{Fig_7}c, square pyramid, IS$_2$ in Fig.~\ref{Fig_6}), which nonetheless fails to fully lock-in. 

This is followed by IS$_3$, containing a distorted, trigonal bi-pyramidal motif as coordination polyhedron; and by IS$_4$ (Fig.~\ref{Fig_7} e), in which a sixth-atom has entered the coordination sphere. IS$_5$ corresponds to $\beta$-tin  (metastep 2660, Fig.~\ref{Fig_7} f), and occurs at a few other metasteps, distinguishable by the characteristic 5-ring number of 136 (Fig.~\ref{Fig_7} main panel, green line). However, the signature on enthalpy is never exactly repeated and remains narrow (Fig.~\ref{Fig_7} main panel, red line, metasteps 1820, 2028, 2660, 3040). While thermodynamics is not favourable to its formation in this regime, its motif is nonetheless part of the landscape, in which it occurs as an intermediate towards other structures, including bct-5, a fact that was already noticed at higher temperature (see discussion above).

\begin{figure}
\begin{center}
\includegraphics[width=0.75\textwidth,keepaspectratio]{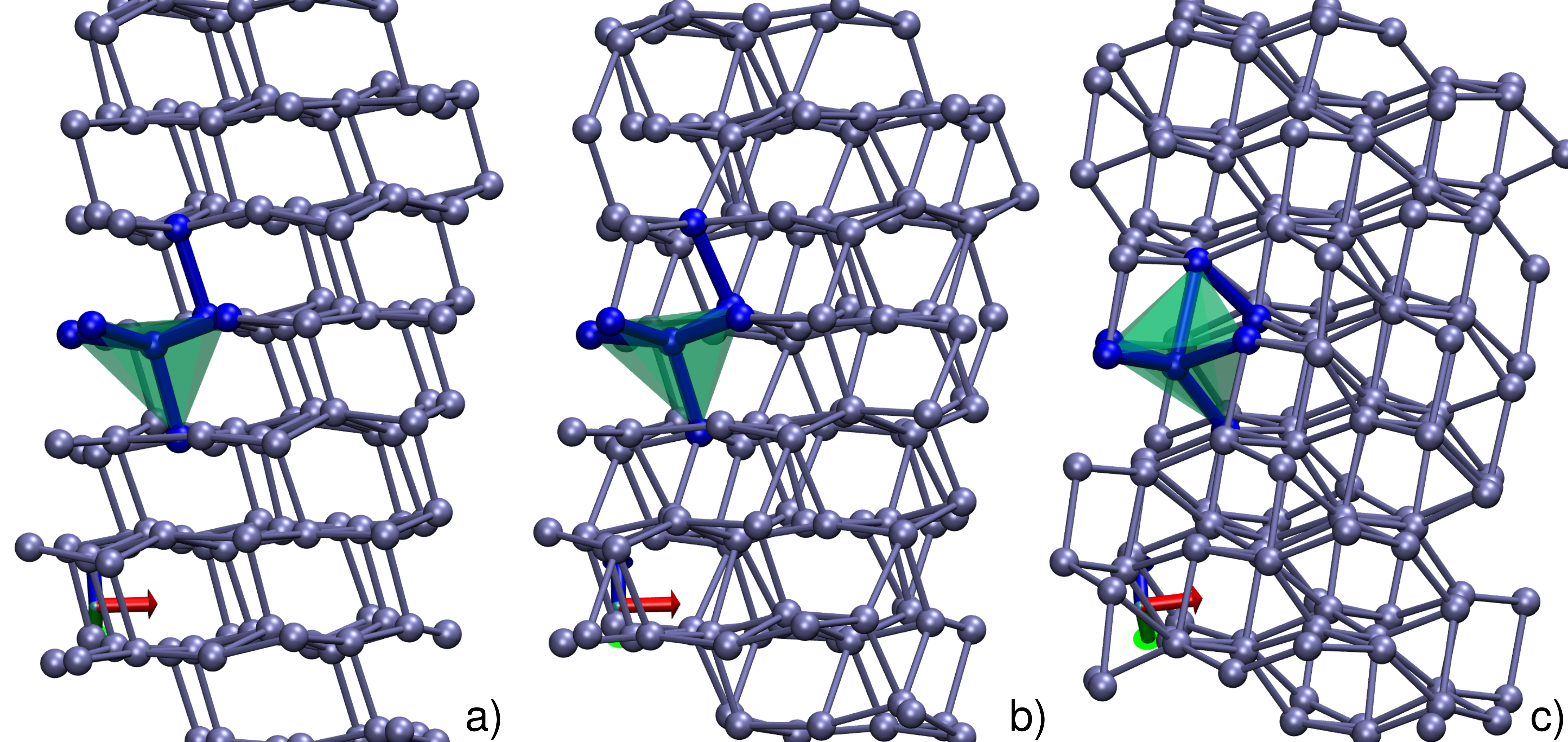}
\end{center}
\caption{Three snapshots from the final part of the metatrajectory of Fig.~\ref{Fig_6} and Fig.~\ref{Fig_7}: bct-5 transforms into a five-connected structural motifs, via a substitution mechanism in the Ge first coordinations sphere. The structural motif of Ge(tI4) is visible in c). The projection was chosen to emphasise the shortening of Ge-Ge contacts caused by this process.}
\label{Fig_8}
\end{figure}

 At metastep 3120 bct-5 locks in, corresponding to full 5-ring depletion and to a plateau of 35 metasteps in the enthalpy, Fig.~\ref{Fig_7} g (its characteristic pyramidal coordination polyhedron is shadowed in green). Shortly thereafter (metastep 3180), bct-5 transforms into a denser (CS {5, 19, 40.5}) motif, containing topologically different Ge atoms, but all 5-fold coordinated. Similar to what was observed above and in a previous work~\cite{Selli:2013coa}, bct-5 is associated with local minima in a region of relative higher enthalpy. 
 
 \section{discussion} 
 
 The modified metadynamics protocol with variable Gaussian shape rescaling, which results into a longer trajectory, allows for many intermediate configurations, some of which are elusive as they result from box shape fluctuations and depend therefore on the specific metadynamics bias history. The enthalpy of Ge(cF136) is higher in any chosen pressure regime. The formation of Ge(tI4), $\beta$-tin succeeded over a direct route for both temperature choices, with normally a sharp lowering of the enthalpy value, a lowering of the ACS, and a persistence of the system in denser configurations, as indicated by the CS (Fig.~\ref{Fig_2} and Fig.~\ref{Fig_5}). Configurationally, Ge$_{136}$ is rigid and requires an extended incubation time for it to transform. This is directly related to the mentioned K$_5$ graphs, which characterise its topology. At higher pressure (Fig.~\ref{Fig_1}), removal of K$_5$ graphs yields shorter ACS and leads to rapid densification. At lower pressures on the contrary, intermediates (Fig.~\ref{Fig_3}) appear, and many amorphous configurations. 
 
 

 At lower temperature and pressure, an intermediate enthalpy regime (Fig.~\ref{Fig_6}) exists, without dominant motifs, but as landscape of configurations rich in five-rings. By inspection of CS and ACS, this region is substantially homogeneous and glass-like. Therein, five-fold rings can occur in different distribution, fused or separated, and are intrinsically flexible, i.e. their configuration can vary. The occurrence of five-rings can therefore be used to measure the distance of a given structural motif from a certain reference structure with less internal degrees of freedom. The enthalpy graph lowers on the average as 5-rings increase (notice that enthalpy and five-ring graphs are arbitrarily aligned in Fig.~\ref{Fig_6}). As well, counting 5-rings provides a stable descriptor to monitor the structural evolution within a rather disordered/amorphous regime.

The occurrence of particular structural motifs can be the result of how Gaussians are accumulated in a metadynamics run, therefore directly interpreting structural sequences into mechanisms can be flawed.  Here we focus just on the last segment of the low-pressure, low-temperature metatrajectory, Fig.~\ref{Fig_6} and Fig.~\ref{Fig_7} (metasteps 3120-3300). Therein, bct-5 rapidly converted into a different 5-connected network (see above), accompanied by enthalpy lowering as the 5-ring numbers increased. The square pyramid around Ge is converted into a distorted trigonal bi-pyramidal shape, by a substitution in the Ge coordination sphere, which generated a set of shorter, parallel contacts visible in the projection of Fig.~\ref{Fig_8} b,c leading to $\beta$-tin.


The competition between denser motifs characterise therefore this metadynamics \textit{megabasin} (1580-3260 metasteps). Importantly, motifs present  at higher pressure, like Ge(tI4), are still operating here, albeit not as lock-in phases, but as transient motifs, particularly around bct-5, to which it can be mechanistically related in this low-pressure, low-temperature regime. 

Based on our detailed mechanistic detailed analysis, and previous results on the role of amorphization~\cite{PhysRevB.105.134107} at higher pressure, a synthetic pathway for bct-5 would therefore require separate steps, including i) the amorphization of the clathrate motif at high pressure followed by ii) recrystallization at a lower pressure such as 2.5 GPa. A possible way to accomplish this protocol might be a short shock wave, which would destabilize the clathrate framework, followed by a low pressure regime, allowing the system to recrystallize.

\section{conclusions}

The compression of Ge$_{136}$ (Ge(cF136)) was studied by means of \textit{ab initio} small-cell metadynamics.  At $p$=5.0 GPa and $T$=300 K, the preferential product was identified as $\beta$-tin (Ge(tI4)). At lower pressure ($p$=2.5 GPa, $T$=300 K)), the fivefold coordinated bct-5 appeared following pronouncedly anisotropic cell fluctuation. To shed further light onto the competition between these two structural motifs, metadynamics runs were also performed at lower temperature $T$= 77 K. $\beta$-tin remained accessible through direct compression at higher pressures, while bct-5 appeared exclusively at lower pressure values. Additionally, lower temperature metadynamics discloses a rich landscape of different, transient five-fold coordinates structural motifs, which can in principle crystallise from amorphous precursors, generated by direct compression of Ge$_{136}$. Different from $\beta$-tin, bct-5 does not contain any five-ring (see Fig.~\ref{Fig_7}), which correlates with its vibrational entropy~\cite{Friedel_1974}, and has a comparatively higher enthalpy/atom. The combination of low pressure, low temperature, non-hydrostatic compression and sufficient time to allow for crystallization can therefore be explored as a gateway for the synthesis of this elusive phase, while higher pressures appear to consistently favour $\beta$-tin (Ge(tI4))~\cite{PhysRevB.105.134107} due to fast crystallization kinetics.

\section{methods}
\subsection{Metadynamics}

Metadynamics~\cite{epjb2011,p13,2006NatMa} allows for the exploration of
the energy surface along one or more collective reaction coordinates. The
method is independent of the level of theory used, it does not require
prior knowledge of the energy landscape and its sampling efficiency can be
enhanced by parallel runs started from different configurations. The
time-evolution of the system is biased by a history-dependent potential, which
discourages the system from visiting already harvested regions of the potential~\cite{Laio:2008iw}. 
Efficiency is achieved in metadynamics also through dimensionality reduction. 
Instead of studying the problem in the full 3N dimensional configuration space of N particles, a relatively small number of collective coordinates
$\mathbf{s}=(s_{1}..s_{m})$ is used instead, which provide a coarse-grained description of the system and are able to distinguish between different free energy minima, i.e. different phases. The inclusion in the space of collective variables of slow degrees of freedom positively impacts the performance of the method.

Each metadynamics metastep consisted of a molecular dynamics run in
the $NVT$ ensemble for a total simulation time of 0.5 ps (timestep 2 fs) at either 300 K or
at liquid nitrogen temperature, 77 K, depending on the regime studied.
All metadynamics runs were performed with 34 atoms in the simulation box which served as a collective (6-dimensional) variable. The size of the minimal box ensured commensurability of all already known phases either open or dense, including Ge(cF136).


\subsection{Density functional computational layers}
SIESTA~\cite{Soler:2002wq} was used as the DFT molecular dynamics layer. For all compression protocols electronic states were expanded by a single-$\zeta$ basis set constituted of numerical orbitals with a norm-conserving Troullier-Martins~\cite{Troullier:1991wi} pseudopotential description of the core electrons. Single-$\zeta$ basis set dramatically reduces computational times providing nonetheless, the right topology and energy differences of all the Ge allotropes under study. The charge density was represented on a real-space grid with an energy cutoff ~\cite{Soler:2002wq} of 200 Ry. A Monkhorst-Pack k-point mesh of 2 $\times$ 2 $\times$ 2 ensured the convergence of the electronic part.  

\subsection{Structure characterization}
To accurately trace structure evolution in the metadynamics simulations,
the average size of the shortest circuits in the structures was calculated~\cite{TOPOS}.
The average circuit size for cF136 is 5.11765, while for $\beta$-tin
it is 4.66667. The average circuit size should decrease along the transition
from the clathrate phase to the $\beta$-tin phase due to a reduction in the number of 5-rings.

To discriminate between $\beta$-tin and bct-5 networks, 
coordination sequences (CS, up to the 3rd shell) were calculated at each metastep.
For $\beta$-tin the CS reads (one atom type): ${6, 22, 46 }$, for bct-5 it is (one atom type): ${5, 16, 33}$. 
Coordination sequences were also used to estimate the percentage of a phase
(either $\beta$-tin or bct-5) along metadynamics run, by looking for atoms with either CS$_{bct-5}$ {5, 16, 33} or CS$_{\beta-tin}$ {6, 22, 46}. 

In case of new structures ideal space group and
asymmetric units were identified with the Gavrog Systre
package~\cite{SYSTRE}.

\section{Acknowledgments}
R.M. was supported by the Slovak Research and Development Agency under Contract No. APVV-15-0496 and by the VEGA Project No. 1/0904/15. S.L. thanks the DFG for support under the priority project SPP 1415 and for a personal Heisenberg fellowship, as well ARCCA Cardiff for the generous allocation of computational resources. He also thanks the Leverhulme Trust for support under Project No. RPG-2020-052. Via our membership of the UK's HPC Materials Chemistry Consortium, which is funded by EPSRC (EP/L000202), this work made use of the facilities of ARCHER, the UK's National High-Performance Computing Service, which is funded by the Office of Science and Technology through EPSRC's High End Computing Programme. We thank Nick Rivier for discussions.

\bibliographystyle{apsrev4-1}
\bibliography{ge}

\end{document}